HOW TO PRINT THIS PAPER:
1. Split this file into a TeX code file + the uuencoded figures file. 
The separation between the files is indicated by the keyword 
"FIGURES" in capitals.
2. Latex the code file (RevTeX 3.0). The macros file "epsf.tex" is
needed for the figures to print from the TeX document.
DELETE UP TO NEXT LINE
\documentstyle[preprint,prb,aps]{revtex}

\begin{document}

%\draft

\title{Mean-field transfer-matrix study of the magnetic phase
diagram of CsNiF$_{\bf 3}$}

\author{Y. Trudeau and M. L. Plumer}

\address{Centre de Recherche en Physique du Solide, D\'epartement de Physique,}
\address{Universit\'e de Sherbrooke, Sherbrooke, Qu\'ebec, J1K 2R1 Canada.}

\date{\today}

\maketitle

\begin{abstract}
A method for treating ferromagnetic chains coupled with
antiferromagnetic interactions on an hexagonal lattice is presented
in this paper. The solution of the $1D$ part of the problem is
obtained by classical transfer-matrix while the coupling between the
chains is processed by mean-field theory. This method is applied with
success to the phase diagram and angular dependence of the critical
field of CsNiF$_3$. Results concerning the general influence of
single-ion anisotropy on the magnetic ordering of such systems are
also presented.
\end{abstract}


% ************************* Introduction ****************************
\section{introduction}

Mean-field theory has long been a useful approximation for the study
of phase transitions in a wide variety of systems, especially
magnetic ones.\cite{stanley} This comes from the great simplification
of dealing with an averaged system instead of explicitely taking into
account each individual interaction. As is generally the case, the
interest here in using mean-field theory is its ability to predict
phase transitions and to follow the evolution of relevant quantities
close to the transition point.  Unfortunately, transition points and
critical exponents extracted from the mean-field theory are found to
be significantly different from the experimental ones, especialy when
the effective dimensionality is small.  This is due to the averaging
process involved which neglects the important critical fluctuations
close to the transition point.

Although mean-field theory has some problems when applied directly to
low dimensionality systems, it can be a very good approximation if
used in conjunction with other techniques. By definition,
quasi-one-dimensional (quasi-$1D$) systems have a direction in which
the energy scale is much larger than in the others. Some numerical
techniques such as classical
transfer-matrix,\cite{blume,loveluck,demers,trudeau2} quantum
transfer-matrix,\cite{delica} and Bethe-Ansatz\cite{takahashi2}
can be used to solve almost exactly the $1D$ part of the problem and
so, instead of using mean-field for all the interactions, it can be
restricted to the coupling between the chains.\cite{scalapino} By
doing this, important fluctuations, although not the critical ones,
will be included and a more accurate solution can be expected.

The present work is mainly concerned with CsNiF$_3$ and, to some extent
other equivalent systems. In this hexagonal insulator, the $S=1$
Ni$^{2+}$ ions are ferromagnetically coupled in chains, along with
the F$^-$ ions.  The chains are well separated
from each other by large Cs$^+$ ions and are coupled by
antiferromagnetic interactions. This arrangment causes a large
spacial anisotropy between the Ni ions, the ratio of the distance
between these ions along the chains and in the basal plane being
0.4205, with an interchain separation of 6.27~$\mathaccent"7017 A$.
Anisotropy is thus present in the magnetic properties, the intrachain
super-exchange ($J_\parallel =$~20~K)\cite{dupas} being much larger
than the interchain one. The exact ratio between the super-exchange
interactions is difficult to estimate because of the strong dipolar
interaction arising between the ferromagnetic chains. It is important
to note that dipolar field is almost nonexistant if the ordering
along the chains is antiferromagnetic (AF), as it is in many other
ABX$_3$ compouds such as CsMnBr$_3$ and CsNiCl$_3$.\cite{diep}

This strong dipolar field is responsible for the particular planar AF
arrangement\cite{plumer1} of CsNiF$_3$ which is different from the
expected 120$^\circ$ structure of a system with only AF interactions
on a triangular lattice. This magnetic phase occurs for temperatures
smaller than $T_N \simeq 2.77$~K.\cite{lussier}.  Lussier {\it et
al.}\cite{lussier} have also shown a strong dependence of the
critical field as a function of the angle between the magnetic field
and the chain axis. As an example, at $T = 2$~K, $H_{c\perp}
\simeq 0.27$~T while for the other direction $H_{c\parallel} \simeq
2.3$~T. An unsuccesful attempt was made to fit this peculiar angular
dependence of the critical field using a mean-field model similar to
the one of Refs.~\onlinecite{plumer1} and \onlinecite{poirier}.  In
this paper, it will be shown that the experimental angular dependence
is typical of a system formed by $XY$ ferromagnetic chains with AF
coupling between them on a triangular lattice.

The Hamiltonian describing the magnetic properties of such systems is
given by:
\begin{equation}
{\cal H } = - J_{\parallel } \sum_{i} {\bf S}_i \cdot {\bf S}_{i+1} - J_{\perp}
\sum_{<a,b>} {\bf S}_{a} \cdot {\bf S}_{b} - g \mu_B {\bf H} \cdot \sum_i {\bf
S}_i + D \sum_i ({\bf S}_i)^2 + {\cal H}_{Dipole}
\label{hamil3d}
\end{equation}
The first, third and fourth sums in equation (\ref{hamil3d}) are
along the chain while the second one is between nearest neighbors in
the plane perpendicular to the chains axis. The strong $XY$ behavior
of CsNiF$_3$ comes from its large single-ion anisotropy ($D$) of
8.5~K.\cite{dupas} The $g$-value was set to 2.2.

As mentioned previously, the usual mean-field theory uses averaged
quantities for the three lattice dimensions. An example of this is
found in Ref.~\onlinecite{plumer1}. The goal of the present work is
to develop a method which treats the chains by classical
tranfer-matrix and uses mean-field for the coupling between them.
This method, refered to as MFTM (Mean-Field Transfer-Matrix), is
presented in section~II. It is followed, in Section~III, by a brief
review of the transfer-matrix algorithm used to calculate the
magnetization and the susceptibility of a $1D$ chain as a function of
temperature and magnetic field. In Section~IV, MFTM is applied to
CsNiF$_3$ for the calculation of its magnetic-field -- temperature
phase diagram and angular dependence of the critical field. Some
other general results, concerning single-ion anisotropy, are
presented in Section~V.

%********************* Champ moyen *******************************

\section{Mean-Field coupling of quasi-1D chains}

\begin{figure}[tpb]
\vglue 0.4cm\epsfxsize 8cm\centerline{\epsfbox{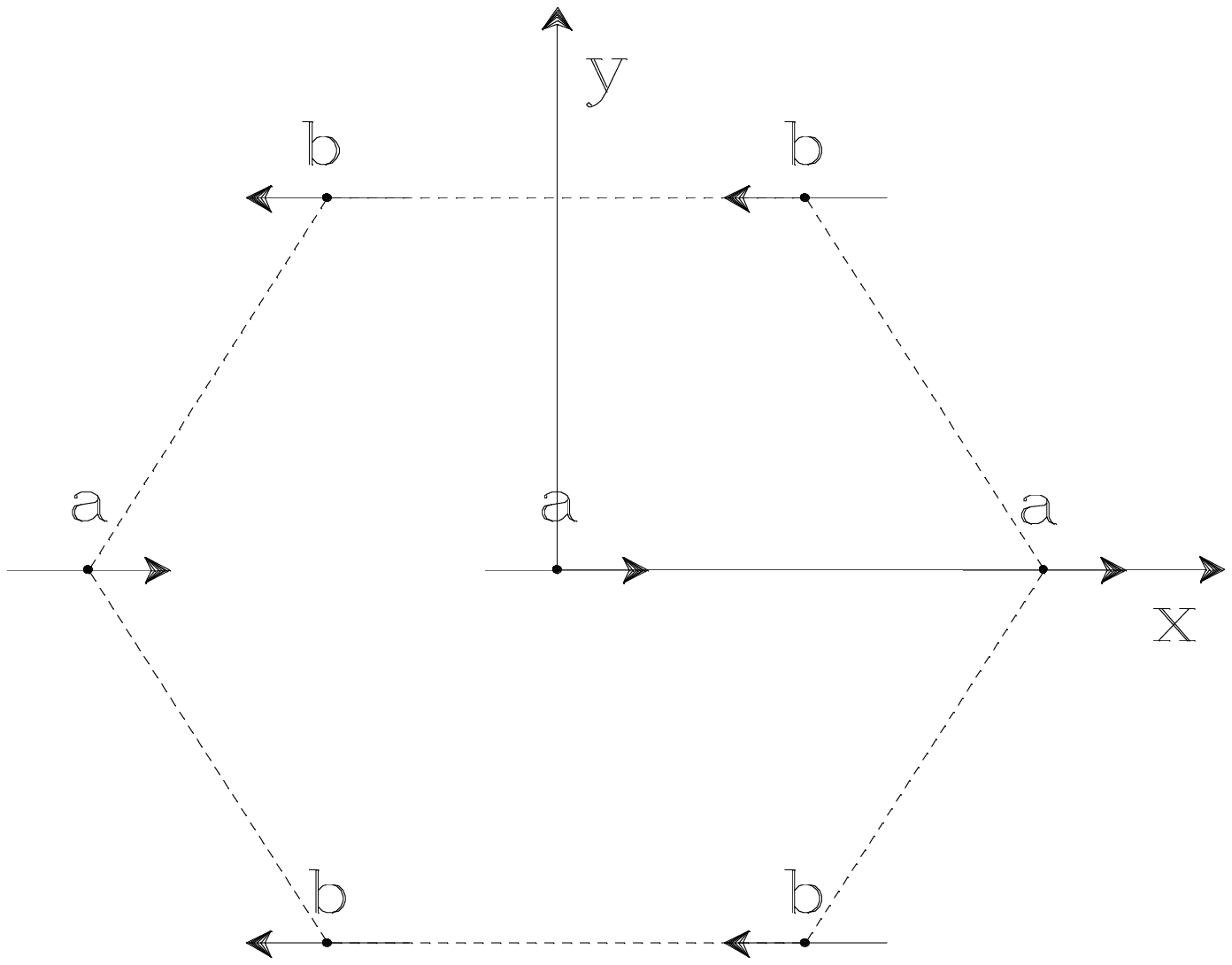}}\vglue 0.4cm
\caption{Planar arrangement of the two magnetic sub-lattices of CsNiF$_3$ for
$H = 0$ and $T < T_c$.}
\label{hexaplan}
\end{figure}

\begin{figure}[tpb]
\vglue 0.4cm\epsfxsize 8cm\centerline{\epsfbox{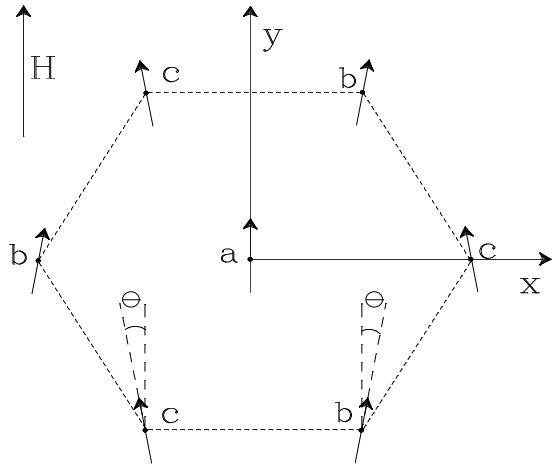}}\vglue 0.4cm
\caption{Ordering of a three antiferromagnetic sub-lattice system at a
magnetic field close to $H_c$.}
\label{120}
\end{figure}

Before presenting the calculated phase diagram of CsNiF$_3$, it
is useful to understand qualitatively the underlying mecanism of the
magnetic order in this system.  For simplicity, consider a starting
point below T$_c$ with a magnetic field oriented in the $XY$ plane,
large enough to destroy the AF order. In this phase,
the paramagnetic one, the free-energy is dominated by a Zeeman term,
$- {\bf m \cdot H}$, and the magnetization ${\bf m}$ is parallel to
the applied field. As the field amplitude is lowered, this term is
diminished up to a point where a perpendicular exchange term,
proportional to $- J_\perp m^2$, becomes equal to it.  At this field,
since $J_\perp < 0$, an AF order develops and a finite angle between
the magnetic sub-lattices appears. The critical field, $H_c$, is
defined where this angle is equal to zero. It is important to notice
the particular planar structure of the magnetic order in CsNiF$_3$
(Fig.~\ref{hexaplan}) which is different from the 120$^\circ$
structure of CsMnBr$_3$, an $XY$ system having strong AF coupling
along the chains. This difference is due to the relatively strong
dipolar field originating from the neighboring ferromagnetic
chains\cite{plumer1}.  This dipolar field depends strongly on the
angle between the magnetic sub-lattices and it can be shown quite
easily that there is no interchain  dipolar field present on a given
site at $H_c$, where the magnetization of all the sub-lattices are
parallel to the field.  So, for small angles between the
sub-lattices, the order starts to develop like a 120$^\circ$ system
(see Fig.~\ref{120}) but at a given angle, the dipolar field coming
from the other chains becomes large enough to flip one of the
sub-lattices according to the planar arrangement of CsNiF$_3$
(Fig.~\ref{hexaplan}).

The starting point of the theoretical description of such a system is
the full three-dimensional ($3D$) Hamiltonian of equation
(\ref{hamil3d}). Since this Hamiltonian, applied to a triangular
lattice, involves up to three magnetic sub-lattices, it is a highly
non-trivial task to solve it even numerically. On the contrary, the
one-dimensional ($1D$) part of the Hamiltonian  consisting of first,
third and fourth term can be calculated numerically quite easily and
with good precision by the tranfer-matrix technique.  The utility of
mean-field theory for this problem comes from the fact that the $1D$
part has a much larger contribution to the free-energy ${\cal F}$
than the $3D$ one ($|J_\parallel| \gg |J_\perp|$).\cite{scalapino}
Thermal averages of the $1D$ Hamiltonian can be used to approximate
the free-energy of the $3D$ one.  Formally, this type of free-energy
is obtained using the Bogoliubov inequality:\cite{critph}

\begin{equation}
{\cal F} \le {\cal F}_T = {\cal F}_{1D} + {\langle {\cal H - H_{\it 1D}} 
\rangle}_{1D} \ .
\label{bogoliubov}
\end{equation}

With the three magnetic sub-lattices of Fig.~\ref{120}, and with
a proper counting of each link in the $J_\perp$ term, this gives the
following trial free-energy:
\begin{eqnarray}
{\cal F}_T = {\cal F}_{1D} &-& J_{\perp} \sum_i  (\ 3 { \langle {\bf
S}_{ia} \rangle }_{1D} \cdot { \langle {\bf S}_{ib} \rangle }_{1D} + 3 {
\langle {\bf S}_{ia} \rangle }_{1D} \cdot { \langle {\bf S}_{ic} \rangle }_{1D}
+ 3 { \langle {\bf S}_{ib} \rangle }_{1D} \cdot { \langle {\bf S}_{ic} \rangle
}_{1D} ) \cr &-& g \mu_B {\bf H} \cdot \sum_i ( { \langle {\bf S}_{ia} \rangle
}_{1D} + { \langle {\bf S}_{ib} \rangle }_{1D} + { \langle {\bf S}_{ic} \rangle
}_{1D}) 
\end{eqnarray}
The thermal averages $\sum_{i} \langle {\bf S}_{ia} \rangle$ are just
the magnetisation ${\bf m}$ of the sublattice $a$. Using the angle
definition of Fig.~\ref{120} for $\theta$ and defining $\phi$ as the
angle between the field and the $XY$ plane, the trial free-energy can
be written as:
\begin{eqnarray}
{\cal F}_T = {\cal F}_{1D} &-& 3 J_{\perp} \ (2 \cos \theta + \cos 2\theta )
m^2 \cr  &-& (1 + 2 \cos \theta) \cos \phi H m_\perp \ ,
\label{ftrial}
\end{eqnarray}
where $m_\perp$ is the component of ${\bf m}$ in the $XY$ plane.

It is also important to note that equation~(\ref{ftrial}) is valid
only for small values of the angle $\theta$ where the dipolar field
has no influence on the order. Minimising (\ref{ftrial}) with respect
to $\theta$ gives $\sin \theta = 0$ or,
\begin{equation}
\cos \theta =  {- 1 \over {2 J_\perp}} \Bigl [ {{\cos \phi H m_\perp} \over
{m^2}} + J_\perp \Bigr ] \ .
\label{cos1}
\end{equation}

This results is reasonable. At high fields, $\theta = 0$,
indicating the field-induced ferromagnetic (paramagnetic) phase.  In
zero field, and for a sufficiently low temperature, one gets the
$120^\circ$ configuration. Of course, this type of structure is not
applicable for CsNiF$_3$ but it reflects the ordering mecanism which
is valid only for small values of $\theta$. As indicated above, $H_c$
is defined where $\cos \theta = 1$ and it can be obtained by solving
self-consistently the following equation:
\begin{equation}
- 3 J_{\perp} m^2 = \cos \phi H m_\perp
\label{self}
\end{equation}
\indent Each side of equation (\ref{self}) represents a contribution
to the free-energy ${\cal F}$ (to within  a minus sign).  The right-hand
side is the magnetic field contribution $- {\bf m} \cdot {\bf H}$,
the dominant one in the paramagnetic phase, while the left-hand is the
$3D$ contribution of the ordered phase. Since ${\bf m}$ is bounded to
unity, the field contribution will eventually dominate at high field
and the phase with the lowest free-energy will be the paramagnetic
one. The strength of the ordered phase is controlled by the parameter
$J_\perp$. In some systems, like CsNiF$_3$, this parameter is not well
known but it can be evaluated with this technique by fitting to
experimental data.

From the limiting behavior of equation~(\ref{self}) at low field, it
is also possible to calculate, within MFTM, the transition
temperature in zero field,\cite{scalapino} $T_c$. This gives:
\begin{equation}
\chi_{1D}(T_c,0) = {{-1} \over {3 J_\perp}} \ ,
\label{chi}
\end{equation}
where $\chi_{1D}$ is the $1D$ susceptibility. An interesting
application of equation~(\ref{chi}) is the possibility to calculate
directly the value of $J_\perp$ needed for a given $T_c$. In the
presents work, this is the only procedure used to set $J_\perp$.

\begin{figure}[tpb]
\vglue 0.4cm\epsfxsize 8cm\centerline{\epsfbox{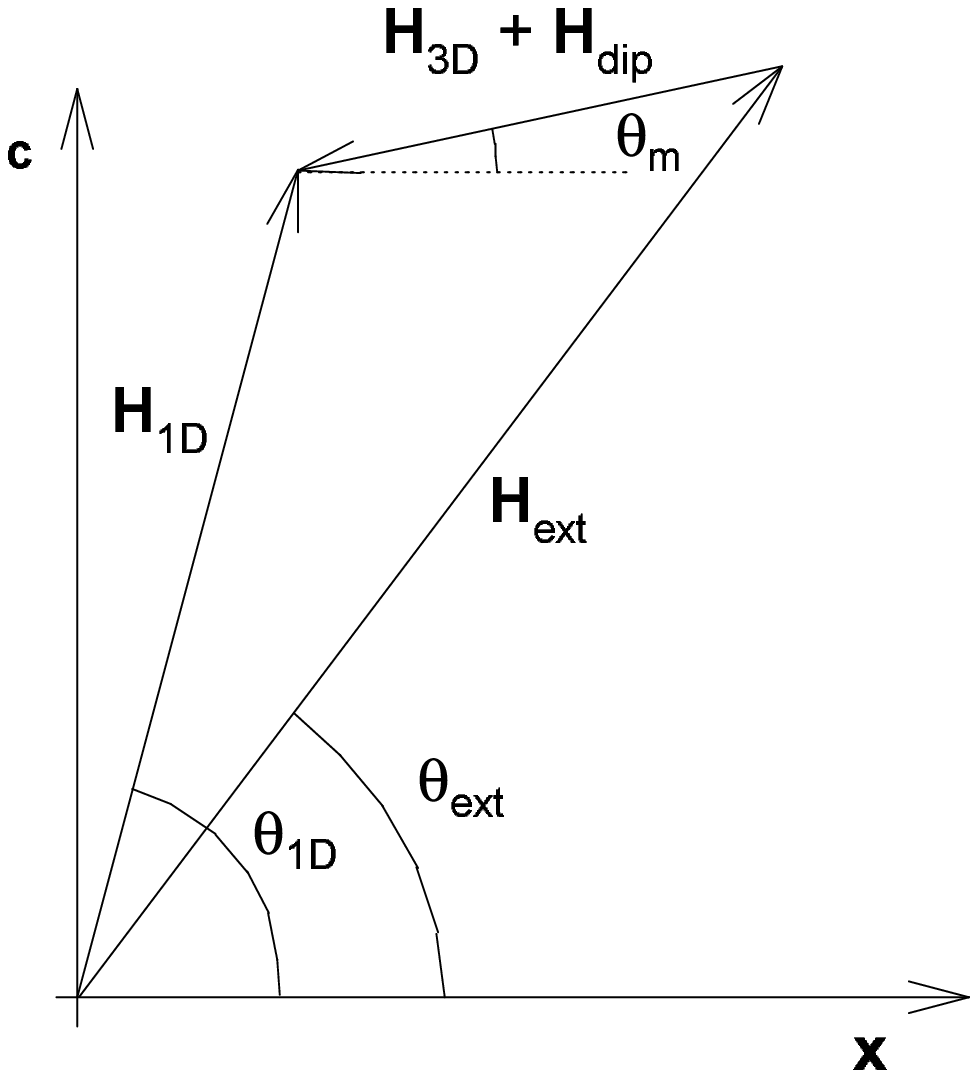}}\vglue 0.4cm
\caption{Vectorial construction giving ${\bf H}_{ext}$}
\label{constH}
\end{figure}

\indent The last item that needs to be solved in this mean-field
theory is the value and the orientation of the external magnetic
field ${\bf H}_{ext}$.  Up to now, the field included in the
calculation was the effective field applied to $1D$ chain. This is
not the true external field; one needs to add a dipolar contribution
${\bf H}_{dip}$ and a $3D$ contribution, ${\bf H}_{3D}$, coming from
the influence of the neighboring chains. Using ${\bf H}_{1D}$ to
designate the field used previously in this paper this gives:
\begin{equation}
{\bf H}_{ext} = {\bf H}_{1D} - {\bf  H}_{3D} - {\bf H}_{dip} \ .
\label{hext}
\end{equation}
\noindent Equation~(\ref{hext}) is shown graphically in
Fig.~\ref{constH}. This figure also shows two new angle
definitions between fields and the $XY$ plane: $\theta_{ext}$ for
${\bf H}_{ext}$ and $\theta_m$ for ${\bf H}_{3D}$ and ${\bf
H}_{dip}$. The evaluation of ${\bf H}_{3D}$ is straigh\-forward it is
just the number of near neighbors times $J_\perp$. It gives:
\begin{equation}
{\bf H}_{3D} = 6 J_\perp {\bf m}
\end{equation}
The dipolar field ${\bf H}_{dip}$ can be evaluated easily, if
the magnetization of all the sub-lattices are parallel, since there
is no interchain contribution. For ${\bf m}$ in the plane
perpendicular to the chain this gives:
\begin{equation}
{\bf H}_{dip} = -32.3 K {\bf m} \ ,
\label{hdip}
\end{equation}
with
\begin{equation}
K = {{g^2 \mu_B^2 \mu_o} \over {4 \pi a^3 k_B}} \ .
\label{K}
\end{equation}
Since CsNiF$_3$ has a quite strong $XY$ caracter, the component of
${\bf m}$ in the $XY$ plane was always by far the greatest for all
the angles $\phi$ used in the calculation so that the numerical
constant used in equation~(\ref{hdip}) was good to a few percent in
all cases.

The fields ${\bf H}_{3D}$ and ${\bf H}_{dip}$ are important in
evaluating the external field ${\bf H}_{ext}$ but they are not 
included in the calculation of $H_c$ by equation~(\ref{self}). The
first one, ${\bf H}_{3D}$, is implicitely accounted for in the left-hand
side of equation~(\ref{self}) while the second one, ${\bf H}_{dip}$,
obviously turns with ${\bf m}$ so there is no free-energy changes due
to it.

The phase transitions are obtained by the solution of equation
(\ref{self}) which implies a prior knowledge of ${\bf m}(T,H)$, the
$1D$ magnetization. An effective and flexible way of calculating
${\bf m}$ is the transfer-matrix technique, which depends on the
various $1D$ parameters like the single-ion anisotropy ($D$) and the
parallel exchange interaction ($J_\parallel$).

% *************************** Matrice de transfert *****************
\section{Transfer-Matrix technique}

Since the present work relies heavily on the tranfer-matrix
technique, a brief summary of this method is presented here.  This
study follows the work of Blume {\it et al.}\cite{blume} and assumes
that the classical spin Hamiltonian for an $N$ spins chain of
magnitude $S$ can be written as:
\begin{equation}
{\cal H}_s = - \sum_{i = 1}^N V( \hat S_i , \hat S_{i+1}) ,
\end{equation}
\noindent with
\begin{equation}
V( \hat S_i , \hat S_{i+1}) = \tilde J \hat S_i \cdot \hat S_{i+1} - {{\tilde
D} \over 2} [(\hat S_i^z)^2 + (\hat S_{i+1}^z)^2] + {{\tilde {\bf H}} \over 2}
\cdot (\hat S_{i} + \hat S_{i+1})
\end{equation}
and
\begin{equation}
\tilde J = {{J_{\parallel} S(S+1)} \over {| J_{\parallel} S(S+1) |}} \ \ \
\Bigl\lbrace {}^{\displaystyle \tilde J = 1 \Rightarrow F}_{\displaystyle
\tilde J = -1 \Rightarrow AF}
\end{equation}
\begin{equation}
\tilde D = {{D S(S+1)} \over {| J_{\parallel} S(S+1) |}} \ \ \
\Bigl\lbrace {}^{\displaystyle \tilde D < 0 \Rightarrow Ising}_{\displaystyle
\tilde D > 0 \Rightarrow XY}
\end{equation}
\begin{equation}
\tilde {\bf H} = {{g \mu_B {\bf H} \sqrt {S(S+1)} } \over {| J_{\parallel}
S(S+1) |} } 
\end{equation}
\begin{equation}
\tilde \beta = \beta {| J_{\parallel} S(S+1) |}
\end{equation}

The unit vectors $\hat S_i$ are oriented along the spin direction.
Using these definitions, the partition function can be
written\cite{critph} as a product of Boltzmann factors,
\begin{eqnarray}
Z &=& \sum_{\{s_i\}} \exp \Bigl [ \tilde \beta \sum^{N}_{i=1} V( \hat S_i , 
\hat S_{i+1}) \Bigr ] \nonumber \cr &=& \sum_{\{s_i\}} \prod_{i=1}^{N} 
\exp \Bigl [\tilde \beta  V( \hat S_i , \hat S_{i+1}) \Bigr ] \ .
\end{eqnarray}
\noindent Since $V$ is translationnaly invariant, the term in the
product does not depend on the site chosen and $Z$ is given by:
\begin{equation}
Z = {\rm Tr} K^N
\end{equation}
with
\begin{equation}
K_{\mu\nu} \equiv \exp \Bigl [ \tilde \beta  V( \hat S_i^\mu , \hat
S_{i+1}^\nu) \Bigr ] 
\end{equation}
Since $N$ is a very large number, only the largest eigenvalue $\lambda_0$ of
$K$ will contribute significantly to $Z$.

\indent For classical spins, the transfer-operator $K$ needs to be
mapped  onto some discretisation of the spherical coordinates in
order to evaluate it numerically. Following the procedure found in
Ref.~\onlinecite{trudeau2}, which is suitable for broken rotational
symmetry, one finds the $1D$ magnetization to given by:
\begin{equation}
{\bf m} = \sum_{i=1}^{N_{\theta}} \sum_{j=i}^{N_{\phi}} W^{\theta}_i W^{\phi}_j
 \Psi_0 (\theta_i , \phi_j )\Psi_0 (\theta_i , \phi_j )
[\cos(\theta_i) \hat z + \sin(\theta_i) \cos(\phi_j) \hat x +
\sin(\theta_i)\sin(\phi_j) \hat y ] ,
\label{mag}
\end{equation}
where $\Psi_0$ is the eigenvector of $K$ corresponding to the largest
eigenvalue $\lambda_0$ and, $N_{\theta}$, $N_{\phi}$, $W^\theta_i$,
$W^\phi_i$, are respectively the number of coordinates and the
Legendre weights of the discret spherical coordinates $\theta_i$ and
$\phi_i$. In the absence of a magnetic field, or when the field is
along the chain axis, a simplification using the rotational
invariance of $V$ can be used.\cite{demers} The reduced number of
coordinates for $K$ greatly improves the amount of computing time
needed for a given precision.  Such a simplification can be used to
evaluate the susceptibility $\chi$ of the zero field
equation~(\ref{chi}). Using the results found in
Ref.~\onlinecite{blume} the $1D$ susceptibility can written as:
\begin{eqnarray}
\chi_{1D}^{zz}(q) &=& \tilde \beta \sum_{l=1}^\infty {{\lambda_{00}^2 -
\lambda_{l0}^2}
\over { \lambda_{00}^2 - 2 \lambda_{00} \lambda_{l0}\cos q + \lambda_{l0}^2 }
} \Bigr [ \sum_{i=1}^{N_\theta} W^\theta_i \cos(\theta_i) \Psi_{00} (\theta_i)
\Psi_{l0} (\theta_i) \Bigl ]^2 \cr
\chi_{1D}^{xx}(q) &=& {\tilde \beta \over 2} \sum_{l=1}^\infty
{{\lambda_{00}^2 - \lambda_{l1}^2} \over { \lambda_{00}^2 - 2
\lambda_{00} \lambda_{l1} \cos q + \lambda_{l1}^2 } } \Bigr [
\sum_{i=1}^{N_\theta} W^\theta_i \cos(\theta_i) 
\Psi_{00} (\theta_i) \Psi_{l1} (\theta_i) \Bigl ]^2 \ .
\end{eqnarray}
\indent When using the rotational invariance as a simplification, one
needs to introduce a second index in order to specify the azimuthal
dependence. This second index is implicitely taken into account with
the broken rotational symmetry algorithm. Typical matrix sizes are
$28^2 \times 28^2 $ for the broken rotational symmetry algorithm
while they are only $64 \times 64$ for the rotational invariance one.
In both cases the amount of computing is not excessive, ranging
between seconds and minutes per $(T,H)$ point. Of course this figure
increases rapidly with matrix size.

% ************************* Resultats ************************
\section{Application to C\lowercase{s}N\lowercase{i}F$_3$}

\begin{figure}[tpb]
\vglue 0.4cm\epsfxsize 8cm\centerline{\epsfbox{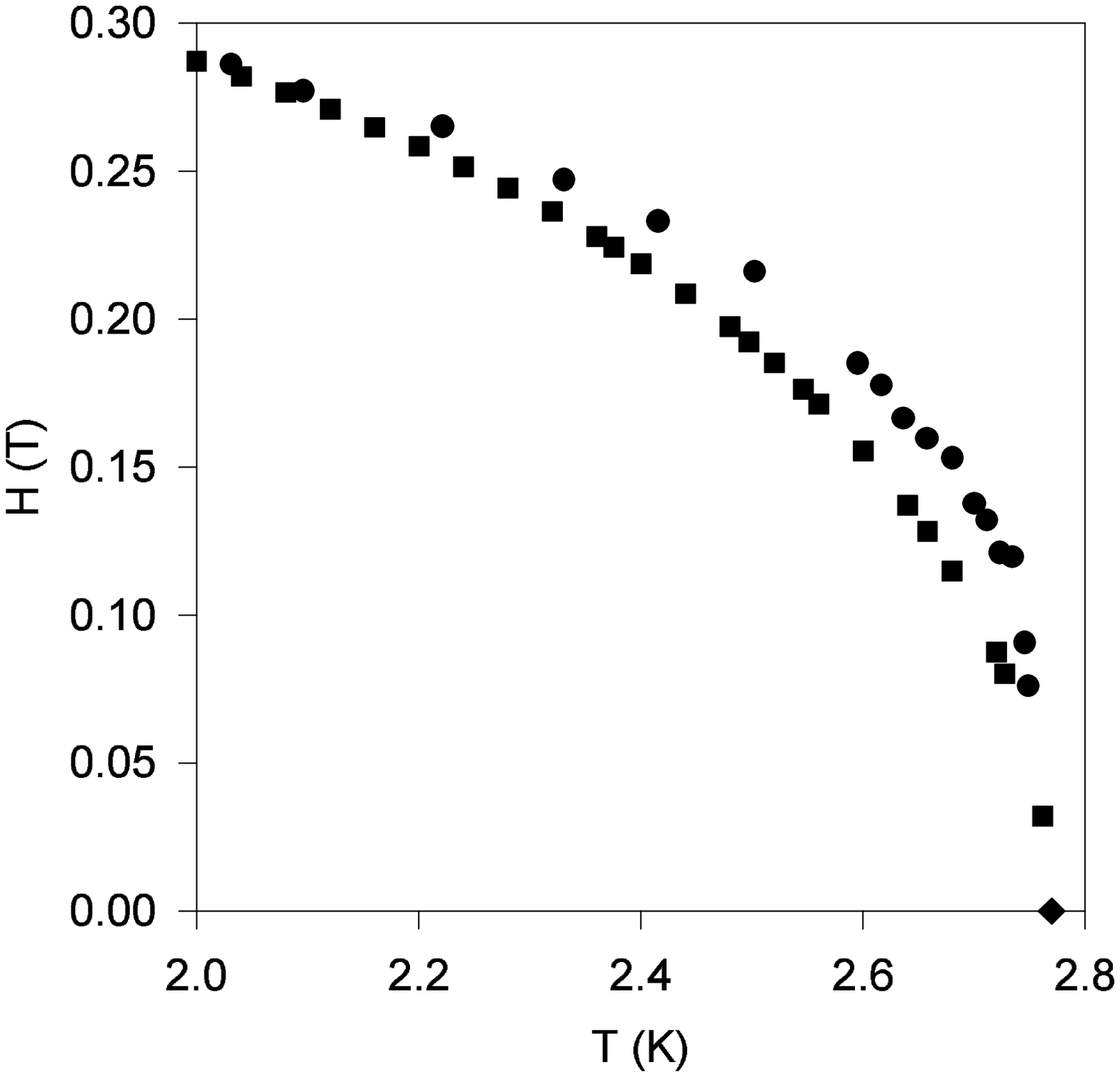}}\vglue 0.4cm
\caption{Comparison between the experimental phase diagram of CsNiF$_3$ given
in Ref.~\protect{\onlinecite{lussier}} (circles) and the one calculated by the
method described in this paper (squares). The diamond represent the zero field
point used to set $J_\perp$ according to eqn.~(\protect{\ref{chi}}).}
\label{diagcsnif3}
\end{figure}

The main goal of this work is to calculate the magnetic phase diagram
of CsNiF$_3$.  Figure~\ref{diagcsnif3} presents a comparison between
the experimental phase diagram obtained with ultrasound by Lussier
{\it et al}.\cite{lussier} and the phase diagram calculated with the
MFTM method. In this figure, the circles are the original
experimental data and the squares are from the MFTM.  The diamond at
zero field is set to 2.77~K by an appropriate choice of $J_\perp$
according to equation~(\ref{chi}). The value found for $J_\perp$
using this procedure is 0.0253~K.  For this phase diagram and all the
subsequent MFTM data on CsNiF$_3$ it will be the only value of
$J_\perp$ used. The overall agreement between theory and experiment
is good considering the fact that this mean-field theory neglects the
$3D$ fluctuations which are not included in the transfer-matrix
calculation and are the relevant ones close to $T_c$. It is quite
obvious from the development of MFTM that $H_c$ scaled as ${\bf m}$
and so $H_c \sim (T_N - T)^\beta$. The absence of the $3D$
fluctuations thus reflects itself in the critical exponent $\beta$,
giving the usual mean-field value of $1/2$ value for the theoretical
phase diagram. This compares with an experimental
estimate\cite{lussier} of $\beta = 0.31 $.

\begin{figure}[tpb]
\vglue 0.4cm\epsfxsize 8cm\centerline{\epsfbox{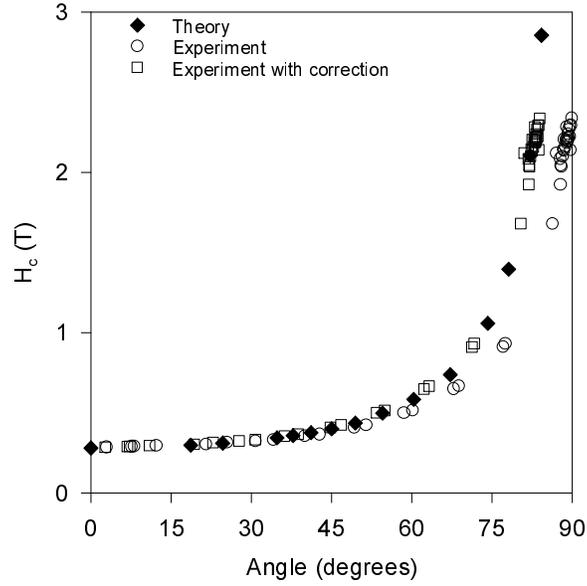}}\vglue 0.4cm
\caption{Comparison between the experimental angular dependence of
$H_c$ at $T = 2$~K given in Ref.~\protect{\onlinecite{lussier}}
(circles and squares) and the one calculated by the method described
in this paper (diamond). The squares represent the original data
(circles) corrected for a plane of rotation misalignment of
6$^\circ$.}
\label{xyvsa}
\end{figure}

The large $XY$ single-ion anisotropy in CsNiF$_3$ has an important
effect on the angular dependence of $H_c$. Crudely, one can say that
when increasing the field for $T \ll T_c$ the transition occurs when
${\bf m}$ saturates so that the left side of equation~(\ref{self})
becomes bounded and the $H$ term at the right can dominate. Because
of the $XY$ anisotropy of CsNiF$_3$, ${\bf m}$ will grow rapidly
along the projection of ${\bf H}$ in the $XY$ plane and it will
saturate for approximately the same field for small angles of the
field out of the plane. For large angles, the projection of ${\bf H}$
in the $XY$ plane will be small and ${\bf m}$ will grow more slowly
giving a larger $H_c$. This type angular dependence is shown in
Fig.~\ref{xyvsa}. In this figure, the experimental data of
Ref.~\onlinecite{lussier} and the MFTM results are represented
respectively by the circles and the diamond. The agreement, already
quite good between these data, becomes excellent when a possible
correction for a misalignement of the experimental plane of rotation
of 6$^\circ$ is included (circles).  This correction, although large,
is not unreasonable. One should also consider the nature of the
theoretical method used as a source of error, especially close to
$\theta = 90$, a pathological angle for the present method since the
coupling between ${\bf m}_\perp$ and $H$ then tends to zero. As
mentioned earlier, even for the largest angle of the magnetic field
out of the $XY$ plane used, ${\bf m}$ was always out the plane by
less than 10$^\circ$ so that the numerical constant in
equation~(\ref{hdip}) was valid to a few percent. The good agreement
found in Fig.~\ref{xyvsa} constrasts sharply with the failure of full
$3D$ mean-field theory, as mentioned in Ref.~\onlinecite{lussier}.

\section{Single-ion Anisotropy}

\begin{figure}[tpb]
\vglue 0.4cm\epsfxsize 8cm\centerline{\epsfbox{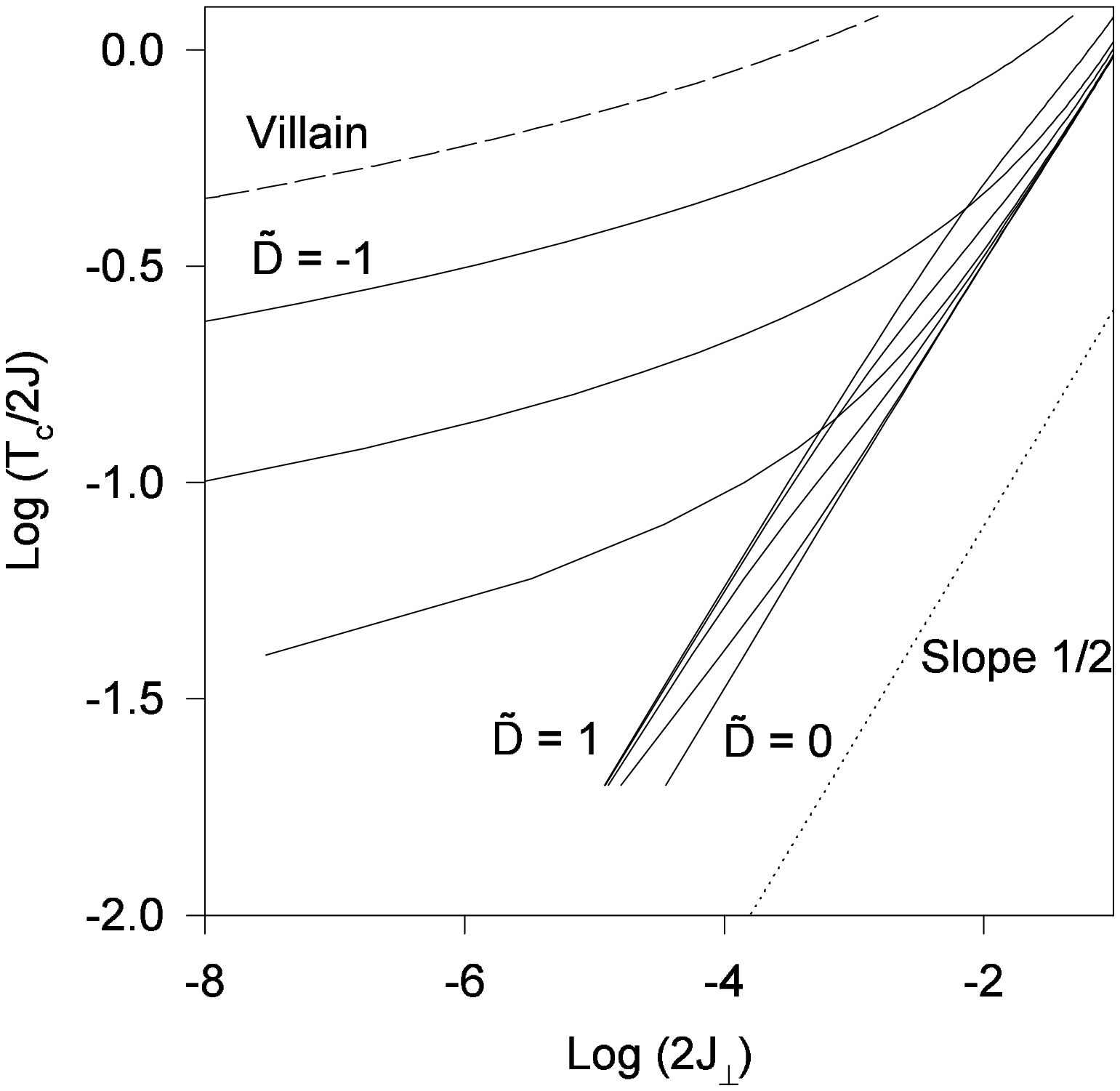}}\vglue 0.4cm
\caption{Logarithm of $T_c$ as a function of the logarithm of $2 J_\perp$.
The curves shown are for D=0, 0.001, 0.01, 0.1, 1, -0.01, -0.1, -1 and the
Villain's expression of Ref.~(\protect{\onlinecite{villain}}).}
\label{tcvsjp}
\end{figure}

More generally, in the absence of a magnetic field, the paramagnetic
phase transitions are determined by the magnetic susceptibility of
the system according to equation (\ref{chi}). When a single-ion
anisotropy is present, like in the Ising and $XY$ cases, the
susceptibility is not isotropic and only the largest component needs
to be taken into account. For an Ising system it is the longitudinal
component $\chi_{1D}^{zz}(0)$ while for an $XY$ one, it is one of the
transverse components $\chi_{1D}^{xx}(0)$ or $\chi_{1D}^{yy}(0)$. The
equation (\ref{chi}) is particularily useful since from the knowledge
of the susceptibility at a given temperature one can extract the
perpendicular exchange interaction $J_\perp$ needed for a phase
transition at this temperature. Such plots, for various values of
$\tilde D$ in a system having $J_\parallel = 1$ and $S = 1$ are shown
in Fig.~\ref{tcvsjp}. One can observe in this figure three limiting
cases; the Heisenberg one at high values of $J_\perp$, the $XY$ one
at intermediate values and the Ising one at low values. This can be
easily understood, the general effect of the single-ion anisotropy is
to favor the order by increasing the susceptibility when reducing the
degrees of freedom of the spin system. It is especially true if the
anisotropy is of the Ising type since the degrees of freedom become
discrete. The $XY$ anistropy has a similar influence but it is less
pronounced, the degrees of freedom being still continuous. Of course,
at high enough temperature $(T \simeq D/k_B)$ the Heisenberg behavior
is recovered in all cases.

There is a functional difference between the continuous and discrete
degrees of freedom. For the continuous ones, the low $J_\perp$
behavior is characterised by a power law of the from $T_c \sim
J_\perp^{1/2}$. The dotted line, having a 1/2 slope, helps appreciate
this. On the contrary, the discrete one (Ising) shows a more complex
exponential form. These results are in agreement with the arguments
of Villain {\it et al}.\cite{villain} Their result for the Ising case
is shown by the dashed curve.

\begin{figure}[tpb]
\vglue 0.4cm\epsfxsize 8cm\centerline{\epsfbox{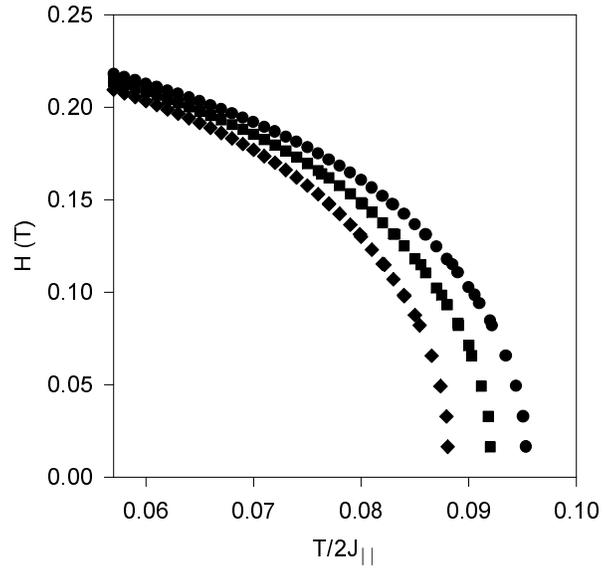}}\vglue 0.4cm
\caption{Magnetic phase diagrams for various D values; 0.5 (circles), 0.2
(squares) and 0.1 (diamond). The $J_\perp$ has been set arbitrarily.}
\label{diagvsd}
\end{figure}

The last aspect investigated in this paper is the influence of the
amplitude of the $XY$ single-ion anisotropy on the whole phase
diagram. As seen in the previous paragraph an increase of the
anisotropy will enhance the magnetic order as a function of the
temperature by increasing the susceptibility. As a function of the
magnetic field, for $T \ll T_c$,  the anisotropy should have a much
smaller effect since it is not implied directly in the phase
transition. Figure~\ref{diagvsd} shows such phase diagrams for a
magnetic field in the $XY$ plane and for three values of $\tilde D$;
0.1, 0.2 and 0.5. The value of $J_\perp$ has been set arbitrarily and
the same lattice parameters than CsNiF$_3$ have been used. In zero
field, the predicted behavior is observed, the highest $T_c$ being
with $\tilde D = 0.5$ while the lowest is with $\tilde D = 0.1$. For
a constant temperature smaller than $T_c$ (like $T/2J_\parallel =
0.06$), one can also observe the small effect of $\tilde D$ on $H_c$.

% *************************** Conclusion ***********************
\section{Conclusion}

In this work, a method for treating antiferromagnetically coupled
ferromagnetic chains by mean-field theory has been developed for a
system with a hexagonal lattice. The solution of the $1D$ part of the
problem has been obtained by a generalised tranfer-matrix algorithm
which is suitable for broken rotational symmetry.\cite{trudeau2}
Although special care has been needed to correctly account for the
large dipolar field originating from the ferromagnetic chains, it has
been possible to reproduce with succes the phase diagram of CsNiF$_3$
by only adjusting the value of $J_\perp$. The peculiar dependence of
the critical field of CsNiF$_3$ as a function of the angle between
the magnetic field and the chain axis has also been reproduced
successfully,\cite{lussier} in sharp contrast with $3D$ mean-field
theory.

More generally, the dependence of the critical temperature as a
function of $J_\perp$ has been established for various values of the
single-ion anisotropy $D$. For the cases having continuous degrees of
freedom, like the $XY$ and Heisenberg ones, a $T_c \sim
J_\perp^{1/2}$ has been found while for the Ising case a more complex
form has been obtained. These results are all in agreement with the
ones obtained by Villain {\it et al} in Ref.~\onlinecite{villain}.
For the $XY$ case, the effect of the size of the single-ion
anisotropy on the overall shape of the phase diagram have also been
investiguated.  The results are in agreement with some simple
physical arguments, an increase of $T_c$ with $D$ and a rather small
effect on $H_c$ for temperatures much lower than $T_c$.

On the basis of the success of this method for the phase diagram of
CsNiF$_3$, some extensions can be considered. The phase diagram of
AF-chains compounds with $XY$ behavior, like CsMnBr$_3$,\cite{diep}
could be examined although some complications with the
transfer-matrix algorithm, coming from the difference in the
effective field of the two staggered sub-lattices, are expected.
Since successive phase transitions occur, the present method, based
on linear response theory, must be generalized so that the
free-energy itself is calculated.\cite{heinonen} A more challenging
prospect is the application of this method to easy-axis AF chains
systems like CsNiCl$_3$.\cite{diep} In this case, the single-ion
anisotropy plays an active role in the magnetic ordering so it cannot
be taken into account only by the transfer-matrix solution of the
$1D$ part of the problem; it must also be included in the calculation
$3D$ free-energy.  This will add at least an order of magnitude to
the complexity of the problem.

% ************************* Remerciements ************************
\acknowledgments
Financial support from the Centre de Recherche en Physique du Solide,
the Natural Sciences and Engineering Research Council of Canada and
le Fonds Formation de Chercheurs et l'Aide \`a la Recherche du
Gouvernement du Qu\'ebec has been essential for this work. We also
want to thank B.~Lussier for his experimental data.

% ************************** References *************************

% ****************** Figure Caption ******************
\input epsf.tex

\end{document}